\documentclass[aps,pre,amsmath,amsfonts,floatfix,superscriptaddress,nofootinbib]{revtex4}

\usepackage{graphicx,graphics}
\usepackage{epstopdf}
\usepackage{amsmath}
\usepackage{rotating}
\usepackage{amssymb}
\usepackage{feynmp}
\usepackage{mathrsfs}
\usepackage{booktabs}
\input{epsf.sty}
\usepackage{bbm}

\def\de{\mathrm d}

\begin{document}
\title{A note on weakly discontinuous dynamical transitions}

\author{Silvio Franz}
\affiliation{Laboratoire de Physique Th\'eorique et Mod\`eles
  Statistiques, CNRS et Universit\'e Paris-Sud 11,
UMR8626, B\^at. 100, 91405 Orsay Cedex, France}

\author{Giorgio Parisi}
\affiliation{Dipartimento di Fisica,
Sapienza Universit\'a di Roma,
INFN, Sezione di Roma I, IPFC -- CNR,
P.le A. Moro 2, I-00185 Roma, Italy
}

\author{Federico Ricci-Tersenghi}
\affiliation{Dipartimento di Fisica,
Sapienza Universit\'a di Roma,
INFN, Sezione di Roma I, IPFC -- CNR,
P.le A. Moro 2, I-00185 Roma, Italy
}

\author{Tommaso Rizzo}
\affiliation{Dipartimento di Fisica,
Sapienza Universit\'a di Roma,
INFN, Sezione di Roma I, IPFC -- CNR,
P.le A. Moro 2, I-00185 Roma, Italy
}

\author{Pierfrancesco Urbani}
\affiliation{Laboratoire de Physique Th\'eorique et Mod\`eles
    Statistiques, CNRS et Universit\'e Paris-Sud 11,
UMR8626, B\^at. 100, 91405 Orsay Cedex, France}
\affiliation{Dipartimento di Fisica,
Sapienza Universit\'a di Roma,
P.le A. Moro 2, I-00185 Roma, Italy
}

\begin{abstract}
  We analyze Mode Coupling discontinuous transition in the limit of
  vanishing discontinuity, approaching the so called ``$A_3$'' point. In these conditions structural relaxation
  and fluctuations appear to have universal form independent from the
  details of the system. The analysis of this limiting case suggests
  new ways for looking at the Mode Coupling equations in the general
  case.
\end{abstract}
\maketitle

\section{Introduction}

The dynamics of supercooled liquids is characterized by a two step
relaxation. After a rapid decay, the dynamical correlation function
displays a plateau where relaxation is arrested before decaying on a
much larger time scale.  Mode Coupling Theory (MCT) describes the
formation of the plateau in terms of a discontinuous dynamical
transition where the length of the plateau diverges as the temperature
become close to the dynamical transition point \cite{Go09}.  

The
approach and the departure from the plateau are described by power laws, respectively $t^{-a}$ and $t^b$, where
the powers $a$ and $b$ are system dependent but obey the universal
relation
\begin{equation}
\lambda=\frac{\Gamma^2(1+b)}{\Gamma(1+2b)}=
\frac{\Gamma^2(1-a)}{\Gamma(1-2a)}\:.\label{l}
\end{equation}
The exponent parameter $\lambda$ also appears in replica theory, where
it has been related to the ratio between six point static correlation
functions that can in principle be measured ore computed directly
using the Boltzmann measure \cite{PR12}. Explicit analytic computations have been
performed in mean-field schematic models \cite{CFLPRR12,CPR12,FLPR12,CFLPR12} and in liquids
\cite{FJPUZ12,Ri12}.  

However, the discontinuous glass transitions is not the
only possibility. A different transition mechanism is found for
example in spin glasses with full replica symmetry breaking, where the
long time limit of the dynamical correlation function passes
continuously from zero to a non zero value when the transition is
crossed.  Within MCT, Bengtzelius, G\"otze and Sj\"olander
\cite{BGS84} have proposed a schematic model whose dynamical
transition can be tuned smoothly from a discontinuous one to a
continuos one through the variation of a parameter.  The resulting
singularity at the continuity point has been named ``$A_3$''.  

The study
of discontinuous/continuous crossover is not a mere academic
exercise.  Realistic systems where this is found include disordered
spin models in presence of a magnetic field, liquids in porous media
both in the MCT \cite{Kr05,Kr07} and in the HNC approximations
\cite{MG88,GS92} and liquid models with pinned particles
\cite{CB12}.  G\"otze and Sj\"ogren \cite{GS99} have studied the scaling 
form of the approach to the plateau of the correlation function
during this crossover within MCT.
However, to the best of our knowledge, a full characterization of the correlation function
in the $\alpha$ regime below the plateau 
has not been presented in the literature.  To fill this gap, we study
the properties of MCT equations for weakly discontinuous
transition. We find that in this case relaxation takes a form which is
universal within the theory. In the way of this taste we find some new
results on schematic MCT equations that have an interest by
themselves. 

In Sec. \ref{Sec:one} we set up the problem and discuss
the dynamical correlation function for weakly discontinuous
transitions both in equilibrium and in the aging regime.  In section
\ref{Sec:two} we extend our analysis to the study of fluctuations and
compute the four point susceptibility. Finally we draw our conclusions.

\section{MCT equations near a continuous transition}\label{Sec:one}
The Mode-Coupling theory postulates that the dynamical correlation and
response functions can be obtained solving a system of
integro-differential equations \cite{Go09}. In the general
theory of liquids these are equations for the dynamical structure
factor and they contain information about the spatial structure of
this quantity. However close to the transition the spatial structure
can be neglected in a first approximation by looking at the peak of
the static structure factor. This has been first underlined by
Bengtzelius \emph{et al.} but it has been noted also in the framework
of the replica approach to the glass transition \cite{FJPUZ12}. Using
this fact one can produce a dynamical equation describing the
evolution of a single mode, that is called the schematic MCT equation.
It is well known that this equation is exactly the one that describes
the Langevin dynamics of fully connected spherical p-spin model with
a 1RSB dynamical transition \cite{CS92,CHS93,BCKM96}.

Approaching
  the dynamical temperature from above we look at the schematic
  Mode-Coupling equation for the correlation function $C(t)$
\begin{equation}
\frac{d C(t)}{dt}=
-TC(t)+(1-C(t))\hat M[C(t)]-\frac 1T\int_{0}^{t}\de u \,\frac{\de C(u)}{\de u} \left(\hat M[C(t-u)]-\hat M[C(t)]\right)\label{MCT}\,,
\end{equation}
where $\hat M[C(t)]$ is the memory kernel that depends on the
parameters of the problem\footnote{For example in the $p$-spin
  spherical model we have $\hat M[q]=pq^{p-1}/2$.}, temperature and/or
density; the initial condition is $C(0)=1$.  Depending on the nature
of the memory kernel, different kind of dynamical transitions are
possible.

The simplest scenario, relevant for supercooled liquids is the one of
a discontinuous transition. Where above and close to the transition
the correlation function display a characteristic two step relaxation
and ergodicity breaking below the transition.

Above the critical point the asymptotic value of the correlation is given by
the unique solution $q_0$ of the equation
\begin{equation}
q=(1-q)M[q].
\label{eqq}
\end{equation}
where we have defined $M[q]=\hat M[q]/T^2$. Close to the dynamical
temperature $T_d$ the correlation function develops a long plateau at
an intermediate level between 1 and $q_0$ before relaxing.  At the
critical temperature the length of the plateau
diverges. Correspondingly, a second solution $q_1>q_0$ to (\ref{eqq})
discontinuously appears. For this value of the correlation one also has
\begin{equation}
1=\frac{d}{dq}(1-q)M[q]|_{q=q_1} 
\label{eqq2}
\end{equation}
which expresses a marginal stability condition of dynamics at
criticality \cite{Go09}.  

Close to the transition, the solution of
  eq. (\ref{MCT}) in the ``alpha regime'', describing the correlation
  decays below $q_1$, verifies the ``time-temperature superposition
  principle'', i.e. it has a scaling form
\begin{equation}
C(t,T)\approx {\bf C}(t/\tau_\alpha(T))
\end{equation}
where the $\tau_\alpha(T)\sim (T-T_d)^{-\gamma}$ is the relaxation
time as a function of the temperature and ${\bf C}(u)$ is a scaling
function independent of the temperature. This scaling function can be
computed from the equation (\ref{MCT}) exactly at the critical
dynamical temperature where one can  neglect the time derivative. The MCT
equation without the time derivative is invariant with respect to
rescaling of time $t\to a t$. Accordingly $\tau_{\alpha}$ is not
directly extracted from this approximation and one can
measure time in arbitrary  units (e.g. in units of $\tau_\alpha$). 

Other kind of transitions are possible if the second solution appears
continuously. As discussed lengthy by G\"otze and collaborators \cite{BGS84}, 
depending on the control parameters in $M$ the transition can change from 
discontinuous to continuous, in passing through a critical point. 

We are interested in the case of weakly discontinuous transitions
close to a critical point where $q_1$ and $q_0$ are almost
degenerate. Thanks to the vicinity to criticality we can characterize
these transitions in a universal way.  For small $q_1-q_0$, the
exponent parameter $\lambda$, which is in general determined by the
relation
\begin{equation}
\lambda=\frac{T_d\hat M''(q_1)}{2(\hat M'(q_1))^{3/2}}\:,\label{lambda}
\end{equation}
is near to 1 and both the exponents 
$a$ and $b$ are close to zero. To the leading order 
$a=b=\sqrt{\frac 6 \pi^2 (1-\lambda)}\sim \sqrt{q_1-q_0} $.
We choose to parameterize the distance from the critical 
point by the value of $b$ itself (so that $q_1-q_0$ is 
a vanishing function of $b$ in the limit $b\to 0$). 

As remarked long ago by G\"otze and Sj\"ogren, at small argument the
function ${\bf C}(t)$ admits a regular short time series expansion in
terms of the parameter $z=t^b$, whose coefficient can be computed
recursively from (\ref{MCT}). 
Unfortunately this expansion is not
convergent in the general case, but for $b\to 0$ we can compute the solution directly 
from the equation. 
More precisely, we suppose the existence of the limit:
\begin{eqnarray}
\lim_{b\to 0; \; t\to\infty\atop z=\left(t/\tau_\alpha \right)^b }({\bf C}(t,b)-q_0)/(q_1-q_0)=G(z)\, ,
\end{eqnarray}
with $G(z)$ a well defined function of its argument. 

Let us now rewrite the equation (\ref{MCT}) in the $\alpha$ regime. We get:
\begin{equation}
C(t)=M[C(t)](1-C(t))-\int_{0}^{t}\de u \,\frac{\de C(u)}{\de u} \left( M[C(t-u)]- M[C(t)]\right)\ .
\label{MCTalpha}
\end{equation}
We now consider the various terms in the equation (\ref{MCT}). 
We firstly consider the memory term in the integral;  in the $b\to 0$ limit
\begin{gather}
M[C(t-u)]-M[C(t)]\simeq M'[C(t)](C(t-u)-C(t))\\
\nonumber
\simeq b z M'(q_1)(q_1-q_0)
\frac{\de G(z)}{\de z}\ln\left(1-\frac{u}{t}\right)
\end{gather}
In an analogous way we have
\begin{gather}
C'(u)=\simeq\frac{b}{u}z (q_1-q_0)\frac{\de G(z)}{\de z}. 
\end{gather}

Next we observe that generically at the transition point the 
 function $N(C)=-C+(1-C)M(C)$ has a single root in 
$q_0$ and a double root in $q_1$. For small $q_1-q_0$ its form should read
$N(C)=-A(C-q_0)(q_1-C)^2$, where by using the relations (\ref{eqq}), 
(\ref{eqq2}) and (\ref{lambda}) we have $A=\frac{M'(q_1)
(1-\lambda)}{q_1-q_0}.$
It follows that to the leading order 
the mode coupling equation can be rewritten as 
\begin{gather}
M'(q_1)(q_1-q_0)^2[(1-\lambda)G(1-G)^2
-(bz)^2\left[G'(z)\right]^2 \int_0^1\frac{\de u}{u}\ln(1-u)] \, .
\label{bzero}
\end{gather}
Now, taking into account that $1-\lambda=b^2\int_0^1\frac{\de
  u}{u}\ln(1-u)]=b^2\frac{\pi^2}{6}$, we obtain the following
 equation for $G$: 
\begin{gather}
G(1-G)^2=z^2\left[G'(z)\right]^2. 
\label{bzero.2}
\end{gather}
This equation is similar to the one found in \cite{GS99} and used there to describe the $\beta$ regime.
Recasting it under the form
\begin{gather}
\frac{dG}{\sqrt{G}(1-G)}=-\frac{dz}{z} \, ,
\label{bzero.3}
\end{gather}
we find that it admits the solutions 
\begin{eqnarray}
G(z)=\left( \frac{1-z/z_0}{1+z/z_0}
\right)^2.
\end{eqnarray}
The value of $z_0$ cannot be computed, as a consequence of scaling
invariance of the MCT equation (\ref{MCTalpha}) and we choose $z_0=1$. We notice that
$G(z)$ decreases from 1 to 0, vanishing at finite $z=z_0$. This is not
in contradiction with the fact that the correlation is positive for
all times at finite $b$, but is a consequence of the fact that we have
taken the limit $b\to 0$.  A detailed computation for small but finite
$b$ tells us that for $z>z_0$ $C(z)\sim e^{-A(z/z_0)^{1/b}}$. This
expression is exponentially small for $b\to 0$ and corresponds to the
simple exponential $C(t)\sim e^{-At/t_0}$ in terms of $t$, where $t_{0}=z_{0}^{1/b}$.

\begin{figure} 
\epsfxsize=250pt \epsffile{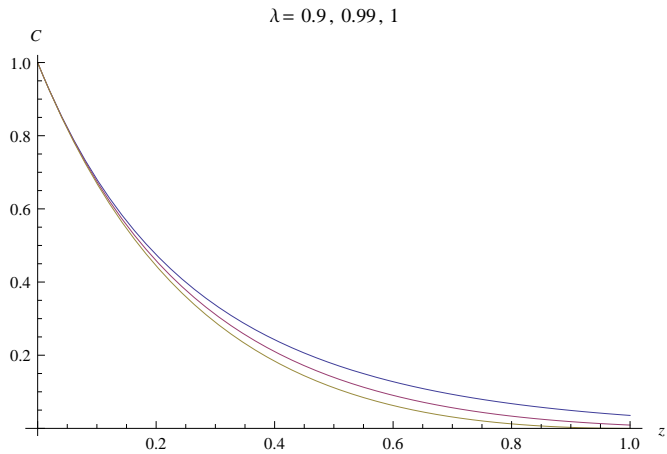}   
\label{fig1}
\caption[0]{Scaling function {\bf C}(z). From top to bottom $\lambda=0.9,\; 0.99,\; 1.$} The first two curves are obtained from the $(20,20)$ Pad\'e approximants
of the small time expansion in $t^b$. The last curve is the function $\left(
\frac{1-z}{1+z}\right)^2$. 
\end{figure}

We can compare this asymptotic solution with the Pad\'e 
approximants of the series expansion of the equation
(\ref{MCT}) for small values of $b$. 
This is done in figure (\ref{fig1}) for the schematic 
$F_{12}$ model \cite{Go09} where $M(C)=\frac{(2 \lambda-1)C+C^2}{\lambda^2}$.
The curves show that the Pad\'e approximants
 give an accurate description of the 
function at  time smaller that 1, and that the limit $\lambda\to 1$ is achieved smoothly.

\subsection{Aging}
The previous analysis can be generalized to the aging dynamics. 
We specialize to the case of the generalized spherical $p$-spin model where the temperature appears explicitly into the equation.
The structure of the equation in the aging alpha regime is similar to the 
equilibrium case and one has \cite{CK93} 
\begin{eqnarray}
&&0=-T C(t,t')+\beta [q_1 f'(q_1)(1-x) - q_0 f'(q_0)x]C(t,t') \label{CUKU} \\
\nonumber
&&+\beta f'(C(t,t'))(1-q_1)-\beta f'(q_1) 
(1-x) C(t,t')\\
\nonumber
&&-\beta x q_0 f'(q_0)+\beta x f'(C(t,t'))(q_1-C(t,t'))\\
\nonumber
&&-\beta x \int_{t'}^t
ds\; \frac{\partial C(t',s)}{\partial s}[f'(C(t,s))-f'(C(t,t'))]\ .
\end{eqnarray}
Here $f'(C)$ generalizes the memory kernel $M$ of the equilibrium
case. 

The quantity $x$ is the so called fluctuation-dissipation
ratio, fixed by the condition that the function 
\begin{eqnarray}
K(C)=-T C+\beta [q_1
f'(q_1)(1-x) - q_0 f'(q_0)x]C +\\ \nonumber
\beta f'(C)(1-q_1)-\beta f'(q_1) (1-x)
C-\beta x q_0 f'(q_0) -\beta x f'(C)(q_1-C)
\end{eqnarray}
 has a double root in
$C=q_1$. 

 It is well known that  equation (\ref{CUKU}) is reparametrization
invariant and admit scaling solutions of the form $C(t,t')={\bf C}
(g(t)-g(t'))$ where the reparametrization function $g(t)$ is left
undetermined. The short time expansion of the equation predicts a
behavior of the kind
\begin{equation}
{\bf C} (u) = q_1 + \left( u \right)^b\ ,
\end{equation}
where $b$ is determined by the condition \cite{BCKM96}
\begin{equation}
\lambda = \frac{T}{2}\frac{f'''(q_1)}{f''(q_1)^{\frac 3 2 }} = x \frac{\Gamma(1+b)^2}{\Gamma(1+2b)}.
\end{equation}

As in the equilibrium case, for $q_1$ close to $q_0$ the function
$K(C)$ behaves as $K(C)=A(C-q_0)(q_1-C)^2)$. We can suppose that
${\bf C}$ becomes an analytic function of 
$z=\left( g(t)-g(t') \right)^b$. 
Notice that if the function $g(t)$ is such that $g''(t)/g'(t)^2 << 1$ for large $t$, then one can equivalently write $z=\left( \frac{t-t'}{\tau_{t'}} \right)^b$.
We can then 
define the scaling function 
\begin{gather} 
G(z)=\lim_{b\to 0, \; t,t'\to\infty \atop z=(g(t)-g(t'))^b}\frac{C(t,t')-q_0}{q_1-q_0}
\end{gather} 
and repeat verbatim the analysis of the equilibrium case. 
It turns out that the equation verified by $G$ coincide with the one found 
at the critical point.  A fortiori, the same is true for the 
function $G(z)$.

\section{Fluctuations}\label{Sec:two} 
In this section we would like exploit our analysis to investigate
fluctuations in the alpha regime.  In the last years, research has
concentrated in the study of fluctuations of the time dependent
correlation functions in terms of 4-point functions. As often in
disordered systems one can define different kinds of correlation
functions with a-priori different scaling properties. It has been
recently proposed that it is useful to disentangle the fluctuations of
correlations with respect to thermal noise for fixed initial condition
from the fluctuations with respect to initial conditions
\cite{FPRR11}.

Denoting by $\langle\cdot\rangle$ the thermal average
for fixed initial condition (iso-configurational average) and by
$[\cdot]$ the average initial condition, we define
\cite{BJ07,FPRR11}
\begin{eqnarray}
\chi_{th}(t)=[\langle C(t)^2\rangle]-[\langle C(t)\rangle^2]\, ,\\
\nonumber
\chi_{het}(t)=[\langle C(t)\rangle^2]-[\langle C(t)\rangle]^2\, .
\end{eqnarray}

A theory for this kind of fluctuations in the beta regime has been
proposed in \cite{FPRR11}, using a "reparametrization invariant"
formulation where time is eliminated in favour of the average
correlation function. Within a gaussian fluctuation theory it is found
that the singularity of $\chi_{het}$ doubles the one of $\chi_{th}$. 

The basic observation allowing to study now the functions in the
$\alpha$ regime is the fact that, as proposed in \cite{BBBKMR07a},
the leading behavior of $\chi_{th}(t)$ can be obtained as
\begin{eqnarray}
\chi_{th}(t)\propto \frac{\partial 
C(t)}{\partial T}. 
\end{eqnarray} 
Before exploiting this relation we would like to
note that it 
appears naturally in the theory put forward in
\cite{FPRR11}. In that context that fluctuations can
be described through a field theory where the correlation function,
which plays the role of fundamental field, couples linearly to the
temperature.  Moreover, the dependence with respect
to the initial configuration turns out to be parameterized by a random
variation of the temperature.  This has the consequence that the
susceptibility $\chi_{het}$ is the square of the thermal one
multiplied by the variance of the random temperature. 

While these consideration strictly hold for the beta and early alpha
regime, the time-temperature superposition principle shows how the
correlation is very sensitive to any temperature change which can
induce large changes in the relaxation time. This is a sort of ``beta
imprinting'' indicating that large fluctuations of the 
correlation fluctuations in the
alpha regime could be just consequence of fluctuations in the initial
time of relaxation.  In last instance this is a consequence of the
emerging scale invariance of the MCT equation when the critical
temperature is approached. We see here a link with the theory of
fluctuations during aging dynamics below $T_d$ developed by
Cugliandolo, Chamon and collaborators \cite{CCC11,CC07,CCCRS04,CCCK02} where
fluctuations are ascribed to the large time emergence of reparametrization
invariance

With all this in mind, we can write:
\begin{eqnarray}
&&N \chi_{th}(t)=\frac{\partial {\bf C}(t/\tau(T))}{\partial T}\ ,\\
\nonumber
&&N \chi_{het}(t)=[\delta T^2]\chi_{th}(t)^2\ . 
\end{eqnarray}
Using the relation $\tau(T)\sim (T-T_d)^{\frac 1 a +\frac 1 b}$ 
with $a \approx b$ for $b\to 0$ and 
${\bf C}(u)=G(u^b)$, 
one gets 
\begin{eqnarray}
&& \chi_{th}(t)=2 \frac{1}{T-T_d}(q_1-q_0) z G'(z) =2 \frac{1}{T-T_d}(q_1-q_0) \sqrt{G}(1-G)\, ,\\
\nonumber
&& \chi_{het}(t)=4 [\delta T^2] \frac{1}{(T-T_d)^2}(q_1-q_0)^2 G(1-G)^2\, .
\end{eqnarray}
The divergence as a function of $T-T_d$ which just depends on 
the power law behavior of the relaxation time, 
confirms the direct dynamical analysis of \cite{BBBKMR07a}. 

Notice that for a finite system the divergence should be cut-off by a
function of the volume. It was found in \cite{SBBB09} and
\cite{FPRR11} that the scaling variable describing
the cross-over is $x=(T-T_d)N^{1/2}$. This predict an alpha relaxation
scaling at $T_d$ where $\chi_{th}\sim \frac{1}{\sqrt{N}}$ and a finite
$\chi_{het}$.

In \cite{FPRR11} it was shown that if $C(t)$ follows a bimodal
distribution as it would be implied by a simple jump process, one
should expect the dependence $\chi_{het} \sim G(1-G)$. Notice the form
we find differ form this expectation. 

We would like to remark that while the square root behavior of
$\chi_{th}$ at small $G$ is only valid in the limit of small $q_1-q_0$
that we are considering, the linear behavior for $G\approx 1$ is more
general: it is a consequence of the initial power law relaxation of
the correlation function $C(t)=q_1-a t^b$, that hold whenever there is
a discontinuous transition.  As far as the small 
$C$ behavior for finite $b$ is concerned, the final
exponential relaxation suggests a behavior $\chi_{th}\sim - C
\log C$.

\section{Conclusions}
The point where the discontinuous transition becomes continuous can be
seen as a critical point for Mode Coupling Theory. As such universal
properties emerge which do not depend of the details of the model
\cite{Go09}. In this note we have computed the scaling
functions for the correlation function both at the MCT transition and
in the aging regime, finding that they take the same universal
form. We have also analyzed the behavior of fluctuations, finding
general expressions of the four point functions as a function of the
correlations.

{\bf Acknowledgments} 

We thank G. Szamel for discussions. SF acknowledges hospitality of the
Physics Department of the ``Sapienza'' University of Rome. The
European Research Council has provided financial support through
European Research Council Grant 247328.

\bibliography{QC}

\end{document}